\newcommand{\be}{\begin{equation}}
\newcommand{\ee}{\end{equation}}
\newcommand{\bea}{\begin{eqnarray}}
\newcommand{\eea}{\end{eqnarray}}
\newcommand{\ba}[1]{\begin{array}{#1}}
\newcommand{\ea}{\end{array}}
\documentclass[12pt]{iopart}
\usepackage{graphicx}
\usepackage{bbm}
\begin{document}
\title{Bright and Dark periods in the Entanglement Dynamics of Interacting Qubits in Contact with the Environment}
\author{Sumanta Das and G. S. Agarwal}
\address{Department of Physics, Oklahoma State University,
Stillwater, OK - 74078, USA} \eads{\mailto{sumanta.das@okstate.edu},
\mailto{agirish@okstate.edu}}
\date{\today}

\begin{abstract}
Interaction among the qubits are basis to many quantum logic operations. We report how such inter-qubit interactions can lead to new features, in the form of bright and dark periods in the entanglement dynamics of two qubits subject to environmental perturbations. These features are seen to be precursors to the well known phenomenon of sudden death of entanglement [Yu $\&$ Eberly, Phys. Rev. Lett. {\bf 93}, 140404 (2004)] for noninteracting qubits. Further we find that the generation of bright and dark periods are generic and occur for wide varieties of the models of environment. We present explicit results for two popular models.
\end{abstract}
\pacs{03.65.Yz, 03.65.Ud, 42.50.Lc}
\maketitle

\noindent{}One of the prime requirements for quantum computation is designing logic gates that can be used to implement algorithms based on the principles of quantum mechanics \cite{niel}. During the last decade substantial theoretical understanding and technological advancement has been acquired in this respect \cite{bennett}.  In analogy to the fundamental gates like XOR in boolean logic, it has been shown that the two qubit logic gate along with single qubit rotation can perform fundamental logic operations (C-NOT) for quantum computation \cite{barenco, vin}. In many cases coherent qubit-qubit interactions has been invoked in constructing such quantum logic gate operations. Moreover such qubit-qubit interactions has become specially important in the context of recent advances in quantum logic gate operations using trapped ions \cite{zoller,wineland} and semiconductor dots \cite{xi, cal,atac,petta}. Note that an important performance factor for a quantum logic gate is its fidelity, which in turn depends on the entanglement among the two qubits. A sustained entanglement among the qubits is a must for optimized gate operation in practical implementation of quantum algorithms. Unfortunately entanglement among quantum systems is extremely fragile and susceptible to decoherence \cite{zurek}, an effect which arises due to unavoidable interaction of the physical system with its environment. Thus study of decoherence effects on the entanglement dynamics and ways to supress it, is of utter importance for quantum information sciences. As an ongoing effort in this respect an important question to study is, how does the inter-qubit interaction in presence of environmental perturbations effect the initial entanglement among the qubits  and thus its operational capability ? In this communication we investigate this important question for two initially entangled interacting qubits.\\
We report that in presence of environmental perturbations the qubit-qubit interaction leads to a new feature in the entanglement dynamics of two qubits. The two initially entangled interacting qubits, get repeatedly disentangled and entangled as they dynamically evolve leading to bright and dark periods in the entanglement. Eventually for longer times we observe ``\textit{entanglement sudden death}" (ESD) \cite{tin}. Note that even though ESD has been studied extensively for non-interacting qubits \cite{lidar, eberly,gordon,almeida, kimble,tso,ficek} in contact with different environments the case of interacting qubit as considered by us is not much studied. An earlier work \cite{hor} had observed and discussed revivals of entanglement due to unitary interactions among the entangled sub-systems. Further it was shown \cite{mani}, that in the strong coupling regime of system-reservoir interaction the deterioration of entanglement can be controlled. In our work we find that the bright and dark periods in the entanglement dynamics are precursor to ESD. Moreover we show explicitly that the phenomenon of generation of bright and dark periods is quite generic and occurs for different kind of models for the environment, like the pure dephasing environment. Further, we also find that this bright and dark periods in entanglement can occur in case of interacting qubits for states which do not exhibit ESD in absence of the interaction. Note that while we concentrate on qubits, Paz and Roncaglia \cite{paz} consider the case of continuous variables i.e harmonic oscillators and demonstrate, in certain parameter domain, such bright and dark periods in entanglement. Our results along with those of ref \cite{paz} would even lead one to think of the existence of such features in entanglement in much larger class of systems. Further, while we focus on the effect of qubit-qubit interactions on deocherence, some recent works have shown how external coherent fields can be used to control the decoherence effects in the entanglement of two qubits \cite{gordon1, fanchini, tahira}.\\
We now discuss our model and show how the interactions between qubits leads to this bright and dark periods in the entanglement. Our model consist of two initially entangled interacting qubits, labeled A and B. Each qubit can be characterized by a  two-level system with an excited state $|e\rangle$ and a ground state $|g\rangle$. Further we assume that the qubits interacts independently with their respective environments. This leads to both local decoherence as well as loss of entanglement of the qubits. The decoherence, for instance can arise due to spontaneous emission from the excited states. Figure 1. show a schematic diagram of our model. 
\begin{figure}[!h]
\begin{center}
\includegraphics[scale = 0.4]{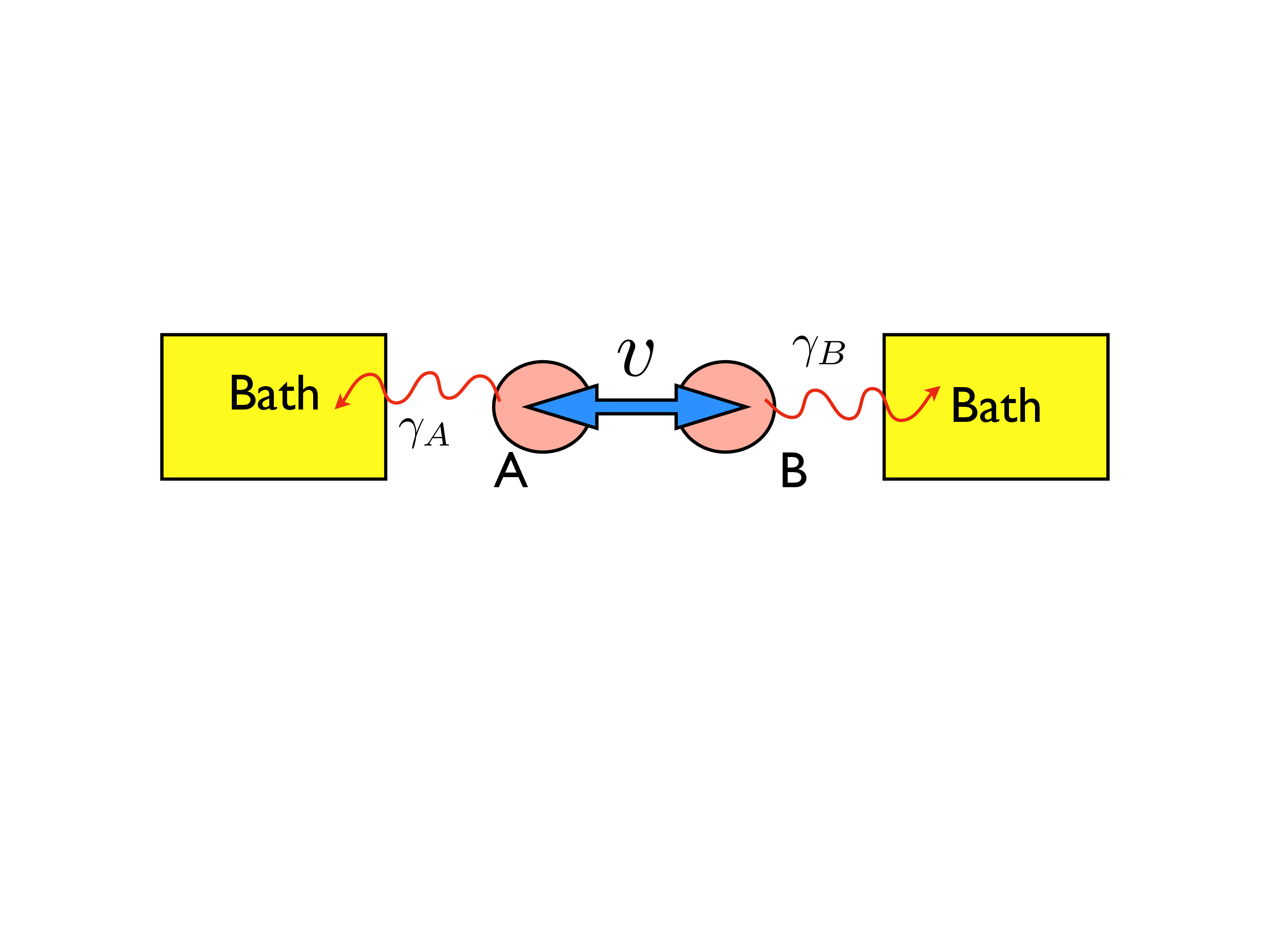}
\caption{Schematic diagram of two qubits modelled as two two-level atom coupled to each other by an interaction parameter $v$. The qubits A and B independently interact with their respective environments (baths) which lead to local decoherence as well as loss in entanglement.}
\end{center}
\end{figure}
The Hamiltonian for our model is then given by,
\be
\label{1}
\mathcal{H} = \hbar \omega_{0}(S^{z}_{A}+S^{z}_{B})+\hbar v(S^{+}_{A}S^{-}_{B}+S^{+}_{B}S^{-}_{A}),
\ee
where $v$ is the interaction between the two qubits and  $S^{z}_{i},S^{+}_{i},S^{-}_{i}$ ($i = $A,B) are the atomic energy, raising and lowering operators respectively which obey angular momentum algebra. We would use the two qubit product basis given by,
\bea
\label{3}
|1\rangle  = |e\rangle_{A}\otimes|e\rangle_{B}&\qquad& |2\rangle  = |e\rangle_{A}\otimes|g\rangle_{B}\nonumber\\
|3\rangle = |g\rangle_{A}\otimes|e\rangle_{B}&\qquad& |4\rangle  = |g\rangle_{A}\otimes|g\rangle_{B}
\eea
Now as each qubit independently interacts with its respective environment, the dynamics of this interaction can be treated in the general framework of master equations. The time evolution of the density operator $\rho$ which gives us information about the dynamics of the system can then be evaluated from the quantum-Liouville equation of motion,
\bea
\label{5}
\dot{\rho}=-\frac{i}{\hbar}[\mathcal{H},\rho]-\sum_{j = A, B}\frac{\gamma_{j}}{2}(S^{+}_{j}S^{-}_{j}\rho-2S^{-}_{j}\rho S^{+}_{j}+\rho S^{+}_{j}S^{-}_{j})\nonumber\\
\eea
where $\gamma_{A}(\gamma_{B} )$ is the spontaneous decay rate of qubit A (B) to the environment.  
To investigate the effect of interaction among the two qubits on decoherence we need to study the dynamics of two qubit entanglement. The entanglement for any bipartite system is best identified by examining the concurrence \cite{wot,buch}, an entanglement measure that relates to the density matrix of the system $\rho$. The concurrence for two qubits is defined as,
\be
\label{7}
C(t) = \max\{0, \sqrt{\lambda_{1}}-\sqrt{\lambda_{2}}-\sqrt{\lambda_{3}}-\sqrt{\lambda_{4}}\},
\ee
where $\lambda$'s are the eigenvalues of the non-hermitian matrix $\rho(t)\tilde{\rho}(t)$ arranged in decreasing order of magnitude. The matrix $\rho(t)$ being the density matrix for the two qubits and the matrix $\tilde{\rho}(t)$ is defined by,
\be
\label{8}
\tilde{\rho}(t)  = (\sigma^{(1)}_{y}\otimes\sigma^{(2)}_{y})\rho^{\ast}(t)(\sigma^{(1)}_{y}\otimes\sigma^{(2)}_{y}),
\ee
where $\rho^{\ast}(t)$ is the complex conjugation of $\rho(t)$ and $\sigma_{y}$ is the usual Pauli matrix expressed in the basis (\ref{3}).
The concurrence varies from $C = 0$ for a separable state to $C = 1$ for a maximally entangled state.
The density matrix needed to evaluate the concurrence for our model should in general have sixteen elements. However, following \cite{tin} we take it as,
\bea
\label{4}
\rho = \frac{1}{3}
\left(\begin{array}{cccc} a & 0 & 0 & 0\\
 0  & b & z& 0\\
 0 & z^{\ast} & c & 0 \\
 0 & 0 & 0 & d\ \end{array}\right),
\eea
where unlike \cite{tin}, we allow the possibility of $z = |z|e^{i\chi}$, to be complex. We have proved from the solution of the quantum-Liouville equation (\ref{5}) that the \textit{initial density matrix} (\ref{4}) \textit{preserves its form for all t}. Finally for our model the concurrence is found to be,
\be
\label{11}
C(t) = \mathsf{Max}\lbrace 0, \tilde{C}(t)\rbrace,
\ee
where $\tilde{C}(t)$ is given by,
\bea
\label{11a}
\tilde{C}(t) & = &2\lbrace|\rho_{23}(t)|-\sqrt{\rho_{11}(t)\rho_{44}(t)}\rbrace.
\eea
Let us now consider a class of mixed states \cite{tin} with a single parameter $a$ satisfying intially $a \geq 0$, $b = c = |z| = 1$ and $d = 1-a$. Then $\rho$ has the form, $\rho\equiv 1/3(a|e_{1}e_{2}\rangle\langle e_{1}e_{2}|+d|g_{1}g_{2}\rangle\langle g_{1}g_{2}|+|\psi\rangle\langle\psi|)$, where $|\psi\rangle\equiv(|e_{1}g_{2}\rangle+e^{i\chi}|g_{1}e_{2}\rangle)$. This has the structure of a Werner state \cite{wer}. The entanglement part of the state depends on $\chi$. Using the solution of (\ref{5}) in (\ref{11a}), we obtain one of our key results
\bea
\label{12}
\tilde{C}(t) & = &\frac{2}{3}e^{-\gamma t}\lbrack(\cos^{2}\chi+\sin^{2}\chi\cos^{2}(2vt))^{1/2}\nonumber\\
&-&\sqrt{a(1-a+2w^{2}+w^{4}a)}\rbrack,
\eea
\begin{figure}
\vspace{0.5in}
\begin{center}
\begin{tabular}{cc}
\includegraphics[scale = 0.5]{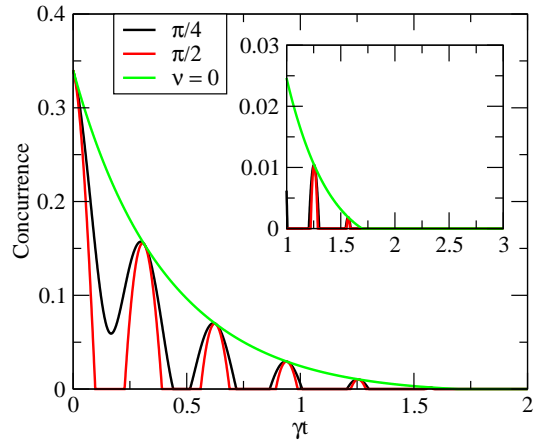}\\
(a)\\
\\
\\
\\
\includegraphics[scale= 0.5]{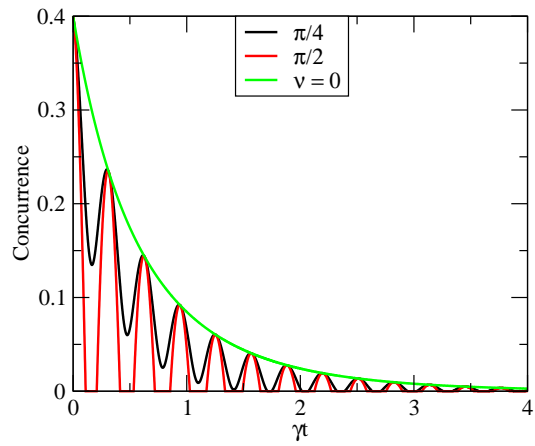}\\
(b) 
\end{tabular}
\caption{Concurrence as a function of time for two initially entangled, interacting qubits with initial conditions $b = c = |z| = 1.0 $ and different initial phases $\chi$. Figures (a) and (b) are for $a = 0.4$ and $a = 0.2$ respectively. The inset in (a) shows the long time behavior of concurrence. The red and black curve in both the figures are for $v = 5\gamma$.}
\end{center}
\end{figure}

where $w = \sqrt{1-e^{-\gamma t}}$. For simplicity we have assumed equal decay rates of both the qubits, $\gamma_{A} = \gamma_{B} = \gamma$. One can clearly see the dependence of $\tilde{C}(t)$ on the interaction $v$ among the qubits and the initial phase $\chi$. We see from (\ref{11}) and (\ref{12}) that in absence of the interaction $v$, concurrence becomes independent of the initial phase and yields the well established result of Yu and Eberly \cite{tin}.\\
Note that $\tilde{C}(t)$ can become negative if,
\bea
\label{13}
 a(1-a+2w^{2}+w^{4}a) > (1-\sin^{2}\chi\sin^{2}(2vt)),
 \eea
in which case concurrence is zero and the qubits get disentangled. To understand how the interaction would effect entanglement we study the analytical result  of equation (\ref{12}) for different values of the parameter $a$  and $\chi$. In figure (2) we show the time dependence of the entanglement for $v = 5\gamma$ and for different values of the initial phase $\chi$. 
The inset of figure 2 (a) shows the long time behavior of entanglement for this case. In figure 2 (a) we show that for $a = 0.4$, the non-interacting qubits ($v = 0$) exhibit sudden death of entanglement (ESD) [visible more clearly in the inset]  whereas when they interact ($v \neq 0$) the concurrence oscillates between zero and non-zero values with diminishing magnitudes and eventually shows ESD. Thus the initially entangled qubits in presence of interaction $v$ gets repeatedly disentangled and entangled before finally becoming completely disentangled. Hence as a result of interaction between the qubits, the concurrence exhibits \textit{bright and dark periods} in the entanglement. Further we observe that when concurrence becomes zero, it remains zero for a time range before reviving. It is worth mentioning here that such bright and dark periodic behavior in entanglement has been predicted for qubits undergoing unitary evolution in a lossless cavity \cite{eberly1}. 
\begin{figure}[!t]
\vspace{0.5in}
\begin{center}
\includegraphics[scale= 0.5]{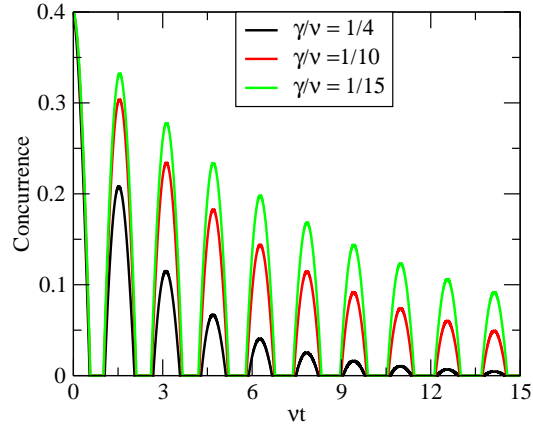}
\caption{Concurrence as a function of time for different decay rates of two initially entangled qubits with initial conditions $a = 0.2 , b = c = |z| = 1, \chi = \pi/2$.}
\end{center}
\end{figure}
This time range is determined by the condition (\ref{13}). In figure 2 (b) we plot the concurrence for $a = 0.2$. Note that for $a=0.2$, no ESD is observed when the qubits are non-interacting and the concurrence monotonically goes to zero as $t \longrightarrow \infty$. For $v \neq 0$, we observe the bright and dark periods in entanglement with diminishing magnitudes and $C(t) \longrightarrow 0$ as $\longrightarrow \infty$.
The figure (3) shows the bright and dark periods in two qubit entanglement for three different  spontaneous decay rates and $a = 0.2$. The initial phase $\chi$ is chosen to be $\pi/2$. For this value of $a$ we observe no ESD but only collapse and revival as expected.\\
\\
{\large{\bf Pure Dephasing of Qubits due to Interaction with the Environment:}}\\
In order to demonstrate the generic nature of our results, we consider other models of the environment. A model which has been successfully used in experiments \cite{kwait} involves pure dephasing. In this case the last two terms in the master equation (\ref{5}) are replaced by,
\bea
\label{14}
-\sum_{i = A,B}\Gamma_{i}(S^{z}_{i}S^{z}_{i}\rho-2S^{z}_{i}\rho S^{z}_{i}+\rho S^{z}_{i}S^{z}_{i})
\eea
where $\Gamma_{A} (\Gamma_{B})$ is the dephasing rate of qubit A (B). Note that in such a model the populations do not decay as a result of the interaction with the environment whereas the coherences  like $\rho_{23}(t)$ decay as $\rho_{23}(0)e^{-(\Gamma_{A}+\Gamma_{B})t}$. Let us now study the the effect of interaction $v$ between the qubits on the dynamics of entanglement. We assume the same initial density matrix of equation (\ref{4}) with the initial conditions $ d = 1-a, b= c = |z| = 1$ and $a \geq 0$ to calculate the concurrence. Under pure dephasing, the \textit{form of matrix in} (\ref{4}) \textit{is preserved for all time}. Using (\ref{11}), (\ref{11a}) and (\ref{14}) we get ,
\bea
\label{17}
\tilde{C}_{D}(t)& = &\frac{2}{3}\lbrack e^{-\tau}\lbrace e^{-2\tau}\cos^{2}\chi+\sin^{2}\chi\lbrace\cos(\Omega^{\prime}\tau)\nonumber\\
& &-\frac{1}{\Omega^{\prime}}\sin(\Omega^{\prime}\tau)\rbrace^{2}\rbrace^{1/2}-\sqrt{a(1-a)}\rbrack,
\eea
 where we assume $\Gamma_{A} = \Gamma_{B} = \Gamma$. Here $\tau = \Gamma t$ and $\Omega^{\prime} = \sqrt{(2v/\Gamma)^{2} -1}$. 
\begin{figure}
\begin{center}
\begin{tabular}{cc}
\includegraphics[scale = 0.5]{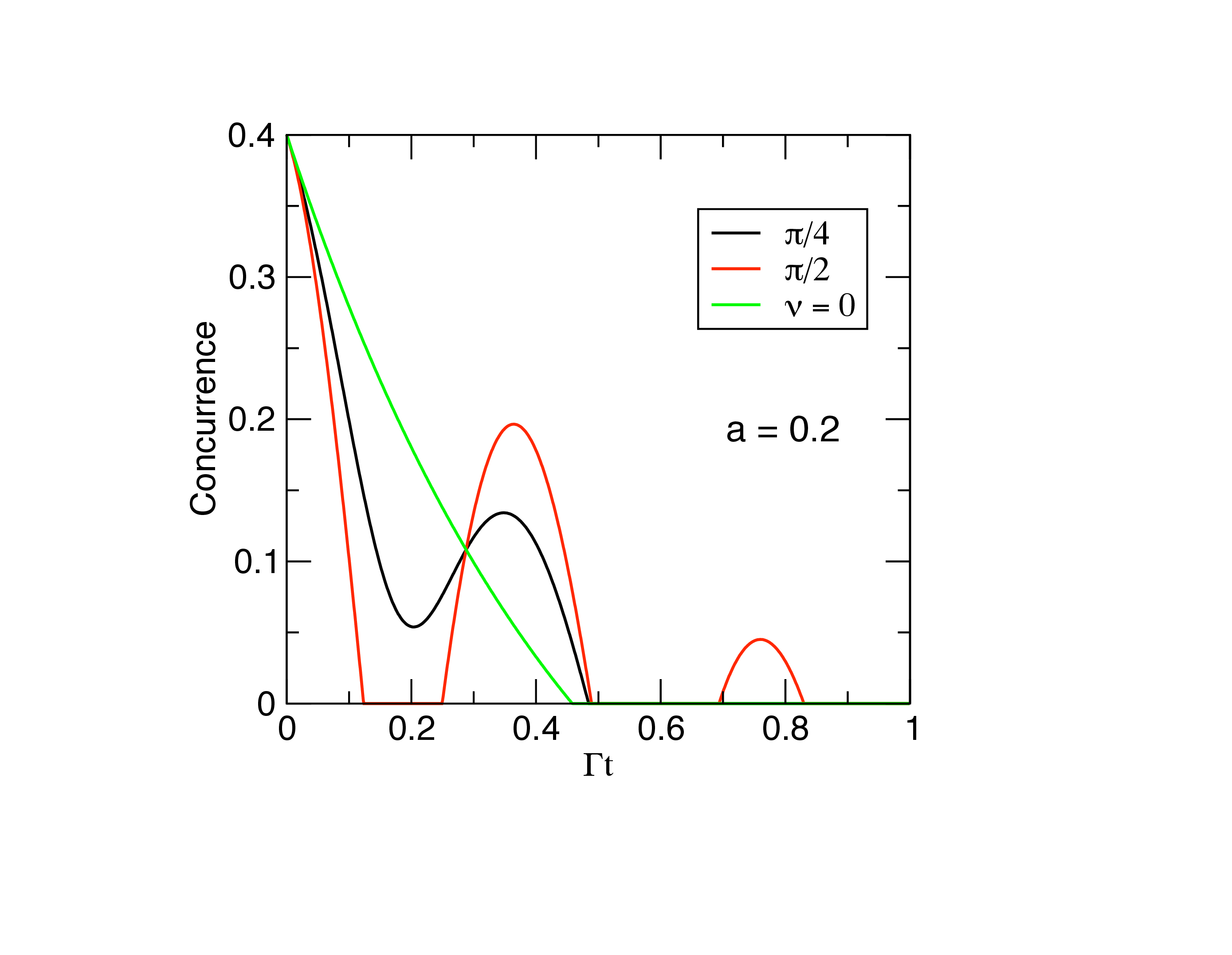}\\
(a)\\
\\
\includegraphics[scale= 0.5]{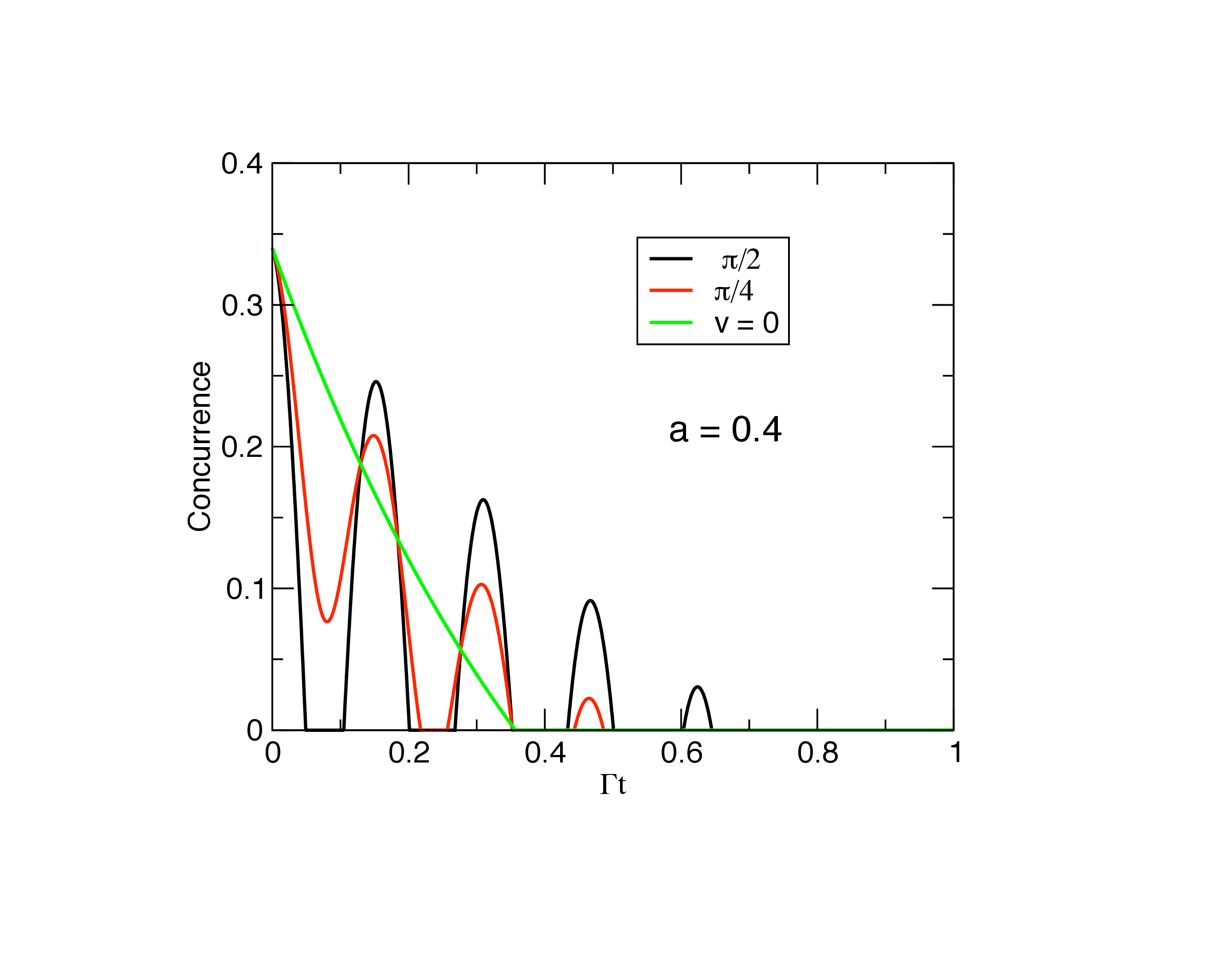}\\
(b)
\end{tabular}
\caption{(Color online) Concurrence as a function of time with initial conditions $ b = c = |z| =1$ and different values of the phase $\chi$ for the dephasing model. The red and black curve in figure (a) if for $v/\Gamma = 4$  and in (b) for $v/\Gamma = 10$.}
\end{center}
\end{figure}
For $v = 0$ we get $\tilde{C}_{D}(t) = 2/3\lbrack e^{-2\tau} -\sqrt{a(1-a)}\rbrack$, which is independent of the initial phase $\chi$. We find \textit{death of entanglement} for $\tau > -1/2\ln \sqrt{a(1-a)}$. Note that Yu and Eberly \cite{eberly} have considered this case earlier but for $a = 1$ only in which case there is no ESD. In figure (4) we show the time dependence of entanglement for a purely dephasing model, for different values of $a$ and initial coherences governed by the phase $\chi$.  From the figures it is seen that  for $v \neq 0$, the two qubit entanglement exhibits the bright and dark periods. Further we also see that for $v \neq 0$, entanglement exhibits this feature even beyond the time when ESD occurs for noninteracting qubits. Moreover figure 4 (b) shows that the frequency of this periodic feature increases with increase in strength of the interaction $v$. The dark period between two consecutive bright periods arises as a result of $\tilde{C}_{D}(t) < 0$, for some time range. This physically means that the two-qubits remain disentangled during this time range. \\
This new feature of bright and dark periods in entanglement should have direct consequences for microscopic systems like ion traps and quantum dots which are currently the forerunner in implementation of quantum logic gates. The interaction between qubits considered in this communication are inherently present in these systems. In quantum dots for example, $\gamma^{-1}\sim$ few ns and one can get a very large range of the parameter $\Gamma^{-1}$ (1-100's of ps) \cite{bori}. Further the interaction strength $v$ can have a range between $1 \mu$ev - $1$ mev depending on gate biasing \cite{atac, petta,taylor}. An earlier study \cite{gert} reports $\gamma \sim 40-100 \mu$ev and coupling strength of $\sim 100-400 \mu$ev, thereby making $v/\gamma \sim 1- 10$ for quantum dot molecules. Thus experimental parameters are in the range we used for our numerical calculation.\\
To summarize we have shown how the interaction between qubits can effect the entangled dynamics of an intially entangled two qubit system. The interaction leads to the formation of bright and dark periods in entanglement. We find this feature for different models of the environment. The frequency of bright and dark periods was found to depend on the strength of interaction between the qubits. Further we find that for noninteracting qubits even when sudden death of entanglement does not occur, entanglement can exhibit this bright and dark periods when the qubits interact. As a future perspective we can investigate the entanglement dynamics for qubits in contact with several different environments. \\
This work was supported by NSF grant no CCF-0829860.


\section*{References}
\begin{thebibliography}{999}

\bibitem{niel}
Nielsen M, and Chuang I, \textit{Quantum Computation and Quantum Information} (Cambridge Univ. Press, Cambridge 2004).

\bibitem{bennett}
DiVincenzo D. P, \textit{Science} {\bf 270}, 255 (1995);
Bennett C. H, and DiVincenzo D. P, \textit{Nature} {\bf 404}, 247 (2000).

\bibitem{barenco}
Barenco A. \textit{et. al.}, Phys. Rev. Lett. {\bf 74}, 4083 (1995).

\bibitem{vin}
Loss D, and DiVincenzo D. P , Phys. Rev. A {\bf 57}, 120 (1998).

\bibitem{zoller}
Cirac J. I, and Zoller P, Phys. Rev. Lett. {\bf 74}, 4091 (1995);
Cirac J. I, and Zoller P, \textit{Nature} {\bf 404}, 579 (2000).

\bibitem{wineland}
DeMarco B \textit{et. al.}, Phys. Rev. Lett. {\bf 89}, 267901 (2002);
Blatt R, and Wineland D. J, \textit{Nature} {\bf 453} 1008 (2008).

\bibitem{xi} 
Li X, \textit{et. al.}, \textit{Science} {\bf 301}, 809 (2003).

\bibitem{cal}
Calarco T, \textit{et. al.}, Phys. Rev. A {\bf 68}, 012310 (2003).

\bibitem{atac}
Imamo$\bar{g}$lu A, \textit{et. al.} Phys. Rev. Lett. {\bf 83}, 4204 (1999);
Robledo L, \textit{et. al.} \textit{Science} {\bf 320}, 772 (2008).

\bibitem{petta}
Petta J. R, \textit{et. al.} \textit{Science} {\bf 309}, 2180 (2005);
Hanson R, and Burkard G, Phys. Rev. Lett. {\bf 98}, 050502 (2007).

\bibitem{zurek}
Zurek W. H, Rev. Mod. Phys. {\bf 75}, 715 (2003).

\bibitem{tin}
Yu T, and Eberly J. H, Phys. Rev. Lett. {\bf 93}, 140404 (2004).

\bibitem{lidar}
Bandyopadhyay S, and Lidar D. A, Phys. Rev. A {\bf 70}, 010301 (2004);
Santos M. F \textit{et. al.}, Phys. Rev. A {\bf 73}, 040305 (2006)

\bibitem{eberly}
Yu T, and Eberly J. H, Phys. Rev. Lett. {\bf 97}, 140403 (2006).

\bibitem{gordon}
Gordon G, and Kurizki G, Phys. Rev. Lett. {\bf 97}, 110503 (2006).

\bibitem{almeida}
Almeida M. P\textit{et. al.}, \textit{Science} {\bf 316}, 579 (2007).

\bibitem{kimble}
Laurat J, \textit{et. al.} Phys. Rev. Lett. {\bf 99}, 180504 (2007).

\bibitem{tso}
Tsomokos D. I, \textit{et. al.}, New. J. Phys. {\bf 9}, 79 (2007).

\bibitem{ficek}
Ficek Z, and Tanas R, Phys. Rev. A {\bf 77}, 054301 (2008).

\bibitem{hor}
Zyczkowski K, \textit{et. al.} Phys. Rev. A {\bf 65}, 012101 (2001).

\bibitem{mani}
Maniscalco S, \textit{et. al.} Phys. Rev. Lett. {\bf 100}, 090503 (2008).

\bibitem{paz}
Paz J. P, and Roncaglia A. J, Phys. Rev. Lett. {\bf 100}, 220401 (2008). This paper also 
contains extensive bibliography on the question of existence of ESD in continuous varibales.

\bibitem{gordon1}
Gordon G, and Kurizki G, Phys. Rev. A {\bf 76}, 042310 (2007);
Gordon G, Euro. Phys. Lett. {\bf 83}, 30009 (2008).

\bibitem{fanchini}
Fanchini F. F, and Napolitano R. d. J, Phys. Rev. A {\bf 76}, 062306 (2007).

\bibitem{tahira}
Tahira R \textit{et. al.}, J. Phys. B {\bf 41}, 205501 (2008).

\bibitem{wot}
Wootters W. K, Phys. Rev. Lett. {\bf 80}, 2245 (1998).

\bibitem{buch}
Walborn S. P, \textit{et. al.}, \textit{Nature} {\bf 440}, 1022 (2006).

\bibitem{wer}
Werner R. F, Phys. Rev. A {\bf 40}, 4277 (1989).

\bibitem{eberly1}
Y$\ddot{o}$gnac M, Yu T, and Eberly J. H, J. Phys. B {\bf 40}, S45 (2007),
A recent paper [Y$\ddot{o}$gnac M, and Eberly J. H, Opt. Lett. {\bf 33}, 270 (2008)] reports such bright and dark periods in entanglement for \textit{noninteracting} qubits driven by single mode quantized fields, which is in a way reminiscent of Jaynes-Cummings dynamics.

\bibitem{kwait}
Kwait P. G\textit{et. al.}, \textit{Science} {\bf 290}, 498 (2000).

\bibitem{bori}
Bori P, \textit{et. al.}, Phys. Rev. Lett. {\bf 87}, 157401 (2001).

\bibitem{taylor}
Krenner H. J, \textit{et. al.}, Phys. Rev. Lett. {\bf 94}, 057402 (2005);
Beirne G, J, \textit{et. al.}, Phys. Rev. Lett. {\bf 96}, 137401 (2006);
Taylor J. M, \textit{et. al.}, Phys. Rev. B {\bf 76}, 035315 (2007).

\bibitem{gert}
Schedelbeck G, \textit{et. al.}, \textit{Science} {\bf 278}, 1792 (1997).


\end{thebibliography}
\end{document}